\documentclass[twocolumn]{aastex701}
\begin{document}

\title{A Catalog of Filaments in the Central Molecular Zone}

\author[0000-0001-8403-8548]{Richard G. Arendt} 
\email{Richard.G.Arendt@nasa.gov}
\affiliation{Center for Space Sciences and Technology, University of Maryland, Baltimore County, Baltimore, MD 21250, USA}
\affiliation{Code 665, NASA/GSFC, 8800 Greenbelt Road, Greenbelt, MD 20771, USA}
\affiliation{Center for Research and Exploration in Space Science and Technology, NASA/GSFC, Greenbelt, MD 20771, USA}

\author[0000-0001-8551-9220]{F. Yusef-Zadeh} 
\email{zadeh@northwestern.edu}
\affiliation{Dept Physics and Astronomy, CIERA, Northwestern University, 2145 Sheridan Road, Evanston, IL 60207, USA}

\author[0000-0001-6864-5057]{I. Heywood} 
\email{Ian.Heywood@skao.int}
\affiliation{SKA Observatory, Jodrell Bank, Lower Whitington, Macclesfield, SK11 9FT, UK}
\affiliation{Department of Physics and Electronics, Rhodes University, PO Box 94, Grahamstown, 6140, South Africa}
\affiliation{South African Radio Astronomy Observatory, Liesbeek House, River Park, Gloucester Road, Cape Town, 7700, South Africa}

\begin{abstract}
We present a catalog of 1.28 GHz radio filaments observed by MeerKAT over the innermost 200~pc 
of the Galaxy (roughly $\pm 1.5\arcdeg$), which includes the central molecular zone (CMZ). 
The catalog is generated by repurposing software developed for the 
automated detection of filaments in solar coronal loops. There are two parts to the catalog. 
The first part, the main catalog, provides a point-by-point listing of locations and basic observational properties 
along each detected filament. The second part is a summary catalog which 
provides a listing of mean, median, or 
total values of various properties for each filament. Tabulated quantities include
position, length, curvature, brightness, and spectral index. 
{The catalogs contain a heterogeneous mix of filamentary structures, including 
nonthermal radio filaments (NRFs), and parts of supernova remnants (SNRs) 
and thermally emitting regions (e.g. \ion{H}{2} regions).}
We discuss criteria for selecting useful subsamples 
of filaments from the catalogs, and some of the details encountered in examining filaments or
selections of filaments from the catalogs.

\end{abstract}


\section{Introduction} \label{sec:intro}

One of the many distinctive features of the Galaxy's central few hundred parsecs
is the presence of a large population of NRFs. 
Since the initial high-resolution observations of the brightest of these in the 
Radio Arc \citep{Yusef-Zadeh:1984},
observations with increasing sensitivity (and resolution) have helped to reveal 
larger numbers of filaments 
\citep{Inoue:1984,Liszt:1985,Haynes:1992,Tsuboi:1995,Yusef-Zadeh:1997,
Staguhn:1998,Lang:1999,LaRosa:2001,LaRosa:2004,Yusef-Zadeh:2004}. 
Most recently, the MeerKAT observations of 
the inner 200 pc of the Galaxy have provided the best view yet \citep{Heywood:2022}, showing that the
filaments run mainly in the direction perpendicular to the Galactic plane and appear to be distributed within 
a radio bubble and an X-ray chimney {\citep{Sofue:1984,Nord:2004,Law:2008,Law:2009,Heywood:2019,Ponti:2019}}. 

While emission of the filaments is clearly nonthermal, as evidenced by their polarization
and spectral index properties
{\citep[e.g.][]{Yusef-Zadeh:1986, Bally:1989, Gray:1991, Yusef-Zadeh:1997,
Pound:2018, Staguhn:2019, Pare:2019, Pare:2021, Pare:2024}}, the origin and nature of the 
filaments remains unclear 
\citep[e.g.][]{Yusef-Zadeh:1984, Serabyn:1994, Shore:1999, LaRosa:2000, Morris:2006, Yusef-Zadeh:2019}.
With the MeerKAT data, we now have a sample of
thousands of filaments from which we can derive properties of the filaments 
as a population (rather than singular examples). In order to do this effectively 
we sought an automated objective means of identifying and characterizing
filaments in the MeerKAT image. We found that software written to detect 
separate filaments in solar coronal loops \citep{Aschwanden:2010}, could be applied successfully 
to detect filaments in this radio mosaic as well. Some of the scientific results 
using this catalog have been presented in \cite{Yusef-Zadeh:2022b,Yusef-Zadeh:2022a,Yusef-Zadeh:2023,Yusef-Zadeh:2024}. 
Here we present the underlying catalog on which those papers were based.
The catalog should be useful for further radio studies, and for detailed 
comparison with observations at other wavelengths.
 
In Section \ref{sec:data} we briefly review the basics of the MeerKAT observations used here.
Section \ref{sec:analysis} describes the application of the automated filament detection, 
and the resulting catalog(s) that are constructed.
Section \ref{sec:selections} provides guidance on possible selection criteria that 
might be used to filter out spurious detections, and subselect among physically 
different types of filaments. Finally in Section \ref{sec:discussion} we discuss
various limitations and properties of the catalog and sources therein.

{Note the catalogs are presented in the measured angular units. To convert 
to physical units, one needs to assume a distance, such as 8.28 kpc 
\citep[for Sgr A*,][]{GRAVITY-Collaboration:2021} which yields a conversion factor
of 0.04 pc asec$^{-1}$. Many objects in the catalog may be much closer 
(or farther) than Sgr A*.}

\section{Data} \label{sec:data}

Here we summarize key aspects of the MeerKAT data, which are
discussed in more detail in \cite{Yusef-Zadeh:2022a} and \cite{Heywood:2022}.
The $L$-band (1.28 GHz) MeerKAT mosaic 
spans $\sim3\arcdeg\times2\arcdeg$ (in Galactic orientation) across the CMZ.
The image itself is in equatorial coordinates with $1.1''$ pixels ($4''$ beam {FWHM}).

To enhance the visibility of filaments, this image 
is filtered using a difference of Gaussians to smooth noise and remove large scale backgrounds.
The is essentially a spatial bandpass filtering of the image.
The smaller, smoothing Gaussian function has $\sigma_1=$ {2.5} pixels. 
The wider, background-subtracting Gaussian function has $\sigma_2=$ {4.5} pixels. 
These values were subjectively determined as ones that helped highlight the 
filamentary features in the image. This process was described in greater 
detail, and illustrated, in \cite{Yusef-Zadeh:2022a}.

We also used a spectral index image constructed from the MeerKAT observations
to characterize the 856 - 1712 MHz spectral index of the identified filaments.
This image formally has lower angular resolution than the total intensity mosaic as it 
is constructed from sub-bands with homogenized resolution, and as such is 
limited by that of the lowest frequency band.

{The total intensity image and the spectral index images from 
\cite{Heywood:2022} are available online.\footnote{\url{https://doi.org/10.48479/fyst-hj47}}}

\section{Analysis} \label{sec:analysis}

Automated identification and tracing of filamentary features in 
the radio image was accomplished using the IDL procedure 
{\tt looptracing\_auto4.pro} in the SolarSoft 
library.\footnote{\url{https://www.lmsal.com/solarsoft/}}
{This code is based on the 
``Oriented Coronal CUrved Loop Tracing'' (OCCULT)
procedure, developed by \cite{Aschwanden:2010} for the purpose 
tracing coronal loops in solar imaging data. The procedure
starts with finding the brightest point in the filtered image.
If this point is identified as part of a ridge structure, then 
the next pixel along the filament
is iteratively identified until an end point is reached where the 
brightness or the change in direction violate preset limits.
Then the filament is similarly delineated in the opposite direction from the 
initial starting point. The two halves are joined, and the emission 
of the filament is subtracted from the image before the procedure repeats
for the next filament. When there are no further sufficiently bright 
filaments, the list of identified filaments is sorted by filament length 
and output as a plain text file.}

Although the 
solar corona and the Galactic center environments are quite different, 
there is a fairly strong similarity in the general morphology of extreme 
ultraviolet imaging of the solar surface and radio imaging of the 
Galactic center. Both images feature filamentary structures, 
often in bundles, superimposed on an irregular background. 

{The adoption of this procedure was motivated by the 
desire for an automated process of identifying NRFs over such a 
large field, and to avoid biases, inconsistencies, and 
inaccuracies that may result 
from manual identification of filaments. 
As such, we did not compare this code against other codes,
although \cite{Aschwanden:2010} had demonstrated that 
OCCULT performed better at tracing solar
coronal loops than several other algorithms.}

\begin{deluxetable*}{lrl}
\label{tab:param}
\tablewidth{0pt}
\tablecaption{{\tt looptracing\_auto4} Control Parameters}
\tablehead{
\colhead{Parameter} &
\colhead{Value} &
\colhead{Description}
}
\startdata
$n_{\rm sm1}$ & 7 & narrow smoothing width [pix]\\
$r_{\rm min}$ & 30.0 & minimum radius of curvature [pix]\\
$l_{\rm min}$ & 30 & minimum length [pix]\\
$n_{\rm struct}$ & 500000 & maximum number of structures to examine\\
$n_{\rm loop}$ & 100000 & maximum number of loops to identify\\
$n_{\rm gap}$ & 2 & maximum length of gaps along a loop [pix]\\
$q_{\rm thresh1}$ & 0.0 & detection threshold for ratio of local image flux to global median flux\\
$q_{\rm thresh2}$ & 5.0 & detection $S/N$ threshold\\
\enddata
\end{deluxetable*}
                               
The input parameters that control {\tt looptracing\_auto4} are listed in Table \ref{tab:param}.
The parameter $n_{\rm sm1}$ sets the scale for smoothing the input image by a $n_{\rm sm1}\times n_{\rm sm1}$ square kernel.
Additionally a version of the image smoothed by a $(n_{\rm sm1}+2)\times(n_{\rm sm1}+2)$ square kernel is subtracted.
Thus this replicates the difference-of-Gaussians filtering that we have already applied to the input image.
Despite this redundancy, the detection results are better than relying solely on the internal smoothing
of the loop tracing code, or than turning off the internal smoothing entirely. 
The parameters $r_{\rm min}$ and $l_{\rm min}$ set the minimum radius of curvature and length of detected filaments. These 
parameters help suppress the detection of spurious background features. The parameters $n_{\rm struct}$ and $n_{\rm loop}$
are used to allocate space for (and thus limit the number of) structures that are examined in the image, and 
subsequent loops that are detected. 
The parameter $n_{\rm gap}$ sets the maximum gap (in pixels) that can occur within 
a single loop. A larger value of $n_{\rm gap}$ can help keep low signal-to-noise ratio ($S/N$) filaments from being split into multiple components,
but may also lead to merging of independent filaments in crowded areas. The parameter $q_{\rm thresh1}$ can
effectively set a floor on the brightness that is recognized as being part of a loop. Given that large scale emission 
is already filtered out of the image, we choose to leave that floor at 0. The parameter $q_{\rm thresh2}$ is used to
stop the searching process when the peak brightness is less than $q_{\rm thresh2}$ times the median of the positive 
intensities in the image.

{Note that all the control parameters relate to the appearance of features in the 
image, not actual physical parameters of the sources. Thus, for images where 
the synthesized beam is well-sampled by the pixel scale, the $S/N$ is reasonably good,  
and the filaments are assumed to be unresolved in width, there should be no need
for major changes in the control parameters. The values listed in Table \ref{tab:param} are 
close to, or the same as, values suggested in \cite{Aschwanden:2010} and the 
documentation for {\tt looptracing\_auto4}. }

\subsection{Main Catalog}

For the Galactic center image and the control parameters listed in Table 1, the code terminates upon 
reaching the $q_{thresh2}$ $S/N$ limit. {\tt looptracing\_auto4} reports $\{x,y\}$ positions, brightness, 
and accumulated length of each detected loop, on a point-by-point basis along each loop. 

After loops were detected, we translated the $\{x,y\}$ positions
into $\{l,b\}$ coordinates, and additionally measured the width of the filament at each point along each loop 
by fitting a Gaussian function perpendicular to the filament. 
The spectral index for each point along each loop is selected from the spectral index map. 
We also calculate a radius of curvature for each point along a filament, as described in the Appendix. 
These parameters are added to the main catalog. 
A sample of the main catalog is presented in Table \ref{tab:main}.

{The intensities in Table \ref{tab:main} match the values in the 
filtered image that is input to {\tt looptracing\_auto4}. However,
because of the mathematical difference-of-Gaussians filtering of the 
input image, these values are artificially low by a factor of 0.1315
as discussed by \cite{Yusef-Zadeh:2022a}.}

\begin{deluxetable*}{rrrrrrrrrrrr}
\label{tab:main}
\tabletypesize{\scriptsize}
\tablewidth{0pt}
\tablecaption{Main Catalog}
\tablehead{
\colhead{ID} &
\colhead{$x$ [pixel]}&
\colhead{$y$ [pixel]}&
\colhead{$l$ [$\arcdeg$]}&
\colhead{$b$ [$\arcdeg$]}&
\colhead{$I_\nu$ [Jy beam$^{-1}$]}&
\colhead{$L$ [pixel]}&
\colhead{$f_g$ [Jy beam$^{-1}$]}&
\colhead{$\delta_g$ [$''$]}&
\colhead{$\sigma_g$ [$''$]}&
\colhead{$\alpha$} &
\colhead{$r_0$ [$''$]}
}
\startdata
1300 & 7418.73 & 4823.88 & -0.748755 & 0.425830 & -3.01443e-06 & 0.0000 & 8.55909e-05 &  4.74470e-01 & 5.99353e+00 & NaN & 2.07258e+02\\
1300 & 7419.68 & 4824.19 & -0.748824 & 0.426127 &  7.98833e-07 & 1.0000 & 2.64174e-05 & -1.51206e-01 & 3.51828e+00 & NaN & 2.07258e+02\\
1300 & 7420.63 & 4824.50 & -0.748894 & 0.426425 &  3.29986e-06 & 2.0000 & 3.58601e-05 & -1.94989e-01 & 4.23308e+00 & NaN & 2.07258e+02\\
1300 & 7421.58 & 4824.80 & -0.748966 & 0.426720 &  4.33616e-06 & 3.0000 & 3.77579e-05 & -5.01826e-01 & 4.71061e+00 & NaN & 2.07258e+02\\
1300 & 7422.53 & 4825.10 & -0.749038 & 0.427016 &  4.16754e-06 & 4.0000 & 4.02011e-05 & -4.31413e-01 & 5.20576e+00 & NaN & 2.07258e+02\\
1300 & 7423.49 & 4825.39 & -0.749115 & 0.427313 &  3.63323e-06 & 5.0000 & 3.92127e-05 & -5.47692e-01 & 5.72810e+00 & NaN & 2.07258e+02\\
1300 & 7424.45 & 4825.68 & -0.749191 & 0.427609 &  3.52501e-06 & 6.0000 & 2.01533e-05 & -5.21050e-01 & 4.88244e+00 & NaN & 2.07258e+02\\
1300 & 7425.41 & 4825.96 & -0.749270 & 0.427904 &  3.95167e-06 & 7.0000 & 1.66714e-05 & -3.33000e-01 & 4.95081e+00 & NaN & 2.07258e+02\\
1300 & 7426.37 & 4826.23 & -0.749352 & 0.428198 &  4.47493e-06 & 8.0000 & 9.45049e-06 & -3.03082e-01 & 3.55585e+00 & NaN & 2.33451e+02\\
1300 & 7427.33 & 4826.50 & -0.749433 & 0.428491 &  4.84325e-06 & 9.0000 & 8.85670e-06 & -5.04256e-01 & 2.79580e+00 & NaN & 2.53935e+02\\
\enddata
\digitalasset
\tablecomments{Table \ref{tab:main} is published in its entirety in the machine-readable format. Only the first 10 lines are shown here. 
There is one line per point along each filament.\\
ID = loop number assigned by {\tt looptracing\_auto4}\\
$x,y$ = location in original image from {\tt looptracing\_auto4}\\
$l,b$ = Galactic longitude, latitude\\
$I_\nu$ = brightness in filtered image from {\tt looptracing\_auto4} {(as in filtered image)}\\
$L$ = cumulative length from {\tt looptracing\_auto4}\\
$f_g$ = cross-sectional Gaussian fit amplitude {(as in filtered image)}\\
$\delta_g$ = cross-sectional Gaussian fit offset\\
$\sigma_g$ = cross-sectional Gaussian fit sigma\\
$\alpha$ = spectral index\\
$r_0$ = local radius of curvature\\
}
\end{deluxetable*}

\subsection{Summary Catalog}
The main catalog is processed into a summary catalog, which for each loop lists parameters that  
can be expressed as single scalar values (Table \ref{tab:summary}). These include the midpoint location of the loop, 
the median brightness, total length, position angle, radius of curvature, medians of the 
cross-sectional Gaussian fits, and the median and gradient of the spectral index along each filament. 
As the spectral index cannot be determined for many low brightness regions, 
we also track the number of points along the filament where the spectral index is defined. 
This is zero for many of the short faint loops that are detected at high latitudes.

Two methods are used for the overall radius of curvature for each filament. 
The calculation of $r_1$ uses the same 3-point method as the local curvature measurement (see Appendix), 
but using only the midpoint and endpoints of the filament. The second method calculates $1/r_2$ using a least 
squares fit using all the points in the filament to determine the radius 
of curvature. A third curvature listed in the summary catalog, $r_3$, is simply the median
of the local curvature measurements ($r_0$) from the main catalog.

{The brightnesses in the summary catalog (Table \ref{tab:summary}) have been 
corrected for the factor of 0.1315 that is induced by the filtering. Therefore,
these brightnesses can be used directly in computing filament luminosities and 
other quantities (e.g. magnetic fields) that may be estimated from the filament 
brightness \citep[see][]{Yusef-Zadeh:2022a}}.

\begin{deluxetable*}{rrrrrrrrrrrrrrrrrr}
\label{tab:summary}
\tabletypesize{\scriptsize}
\tablewidth{0pt}
\tablecaption{Summary Catalog}
\tablehead{
\colhead{ID} &
\colhead{$x$ [pixel]}&
\colhead{$y$ [pixel]}&
\colhead{$l$ [$\arcdeg$]}&
\colhead{$b$ [$\arcdeg$]}&
\colhead{$I_\nu$ [Jy beam$^{-1}$]}&
\colhead{$L$ [$''$]}&
\colhead{$\theta_{\rm e}$ [$\arcdeg$]}&
\colhead{$\theta_{\rm g}$ [$\arcdeg$]}&
\colhead{$r_1$ [$''$]} & 
\colhead{$r_2$ [$''$]} & 
\colhead{$r_3$ [$''$]} & 
\colhead{$f_{\rm g}$ [Jy beam$^{-1}$]}&
\colhead{$\sigma_{\rm g}$ [$''$]}&
\colhead{$\alpha$} &
\colhead{$\sigma_\alpha$} &
\colhead{$N_\alpha$} &
\colhead{$d\alpha/dL$ [$('')^{-1}$]}
}
\startdata
1300 & 8054.14 & 4861.11 & -0.8397466 &  0.5976747 & 1.881e-04 & 1278.20 &  90.12 & 148.86 & 2999.52 & 10775.85 & 1379.92 & 3.571e-04 & 3.35 & -1.373e+00 & 3.084e-01 & 1021 &  7.941e-05\\
 252 & 3835.32 & 7056.79 &  0.4025384 & -0.1561897 & 7.318e-04 & 1053.80 & 114.41 & 173.16 & 2459.79 &  4759.91 & 1485.45 & 1.270e-03 & 3.38 & -9.145e-01 & 2.224e-01 &  915 & -3.858e-04\\
 986 & 5197.30 & 3147.72 & -0.8344428 & -0.4201037 & 2.925e-04 & 1041.70 & 101.05 & 159.80 & 1613.91 &  2179.08 & 1052.12 & 2.102e-04 & 3.78 & -7.304e-01 & 2.384e-01 &  903 &  4.281e-05\\
1896 & 6162.82 & 6391.13 & -0.1402906 &  0.3462411 & 1.891e-04 &  886.60 & 121.74 &   0.48 & 2240.51 & 18600.35 & 1104.64 & 3.555e-04 & 3.85 & -1.288e+00 & 4.109e-01 &  618 &  2.068e-04\\
  89 & 6277.30 & 5240.94 & -0.4588813 &  0.1938145 & 3.285e-04 &  851.40 &  95.06 & 153.81 & 1385.11 &  1617.66 &  776.64 & 4.654e-04 & 3.08 & -7.369e-01 & 4.055e-01 &  602 & -1.984e-05\\
 452 & 4451.06 & 5379.72 & -0.1331216 & -0.2612211 & 4.475e-04 &  842.60 &  86.24 & 144.99 & 1632.87 &  2452.40 & 1308.00 & 1.314e-03 & 3.61 & -1.022e+00 & 2.774e-01 &  574 &  9.227e-05\\
 852 & 4014.29 & 4379.29 & -0.3252106 & -0.5339017 & 2.993e-04 &  816.20 & 173.09 &  51.84 & 1133.96 &  1387.52 &  896.78 & 5.478e-04 & 3.40 & -3.745e-01 & 2.058e-01 &  651 &  6.239e-06\\
 602 & 4351.01 & 5153.60 & -0.1763268 & -0.3232019 & 3.716e-04 &  788.70 &  93.19 & 151.94 & 1732.22 &  3033.75 & 1345.76 & 8.187e-04 & 3.24 & -7.299e-01 & 3.568e-01 &  641 & -5.705e-04\\
 166 & 5474.54 & 6316.10 & -0.0507785 &  0.1545656 & 1.125e-03 &  782.10 & 107.98 & 166.73 & 1985.36 & 31459.71 & 1344.63 & 3.885e-03 & 4.14 & -7.521e-01 & 2.130e-01 &  582 &  6.508e-04\\
  83 & 5899.82 & 5861.36 & -0.2369799 &  0.1935661 & 8.067e-04 &  764.50 &  76.56 & 135.30 &  951.49 &  1354.98 &  667.39 & 1.833e-03 & 3.34 & -8.048e-01 & 2.702e-01 &  597 &  2.735e-04\\
\enddata
\digitalasset
\tablecomments{Table \ref{tab:summary} is published in its entirety in the machine-readable format. Only the first 10 lines are shown here. 
There is one line per filament.\\
ID = loop number assigned by {\tt looptracing\_auto4}\\
$x,y$ = location of midpoint\\
$l,b$ = Galactic coordinates of midpoint\\
$I_\nu$ = median brightness in filtered image {(corrected for filtering)}\\
$L$ = total length\\
$\theta_{\rm e}$ = position angle (equatorial)\\
$\theta_{\rm g}$ = position angle (Galactic)\\
$r_1$ = radius of curvature (3-point method, center+endpoints)\\
$r_2$ = radius of curvature (full fit) \\
$r_3$ = median local radius of curvature ($r_0$) \\
$f_{\rm g}$ = median cross-sectional Gaussian fit amplitude  {(corrected for filtering)}\\
$\sigma_{\rm g}$ = median cross-sectional Gaussian fit sigma\\
$\alpha$ = median spectral index\\
$\sigma_\alpha$ = standard deviation spectral index\\
$N_\alpha$ = number of pixels with valid spectral index\\
$d\alpha/dL$ = linear gradient of the spectral index\\
}
\end{deluxetable*}

\section{Selections}\label{sec:selections}
The {\tt looptracing\_auto4} code was run to provide automated identification of 
all detectable filaments, with minimal prior constraints on what is deemed to be a 
filament (see Table \ref{tab:param}). In examining the results, we find that some 
clearly spurious features were found along faint arcuate discontinuities 
in the mosaic image where edges of fields overlap. A number (113) of these 
were manually excluded from the final catalog.
It is intended that further decisions on the separation of real filaments in the catalog
from spurious low $S/N$ features and artifacts be done by the user based on the
scientific objectives. 

\begin{figure}[t]
   \centering
   \includegraphics[width=3.35in]{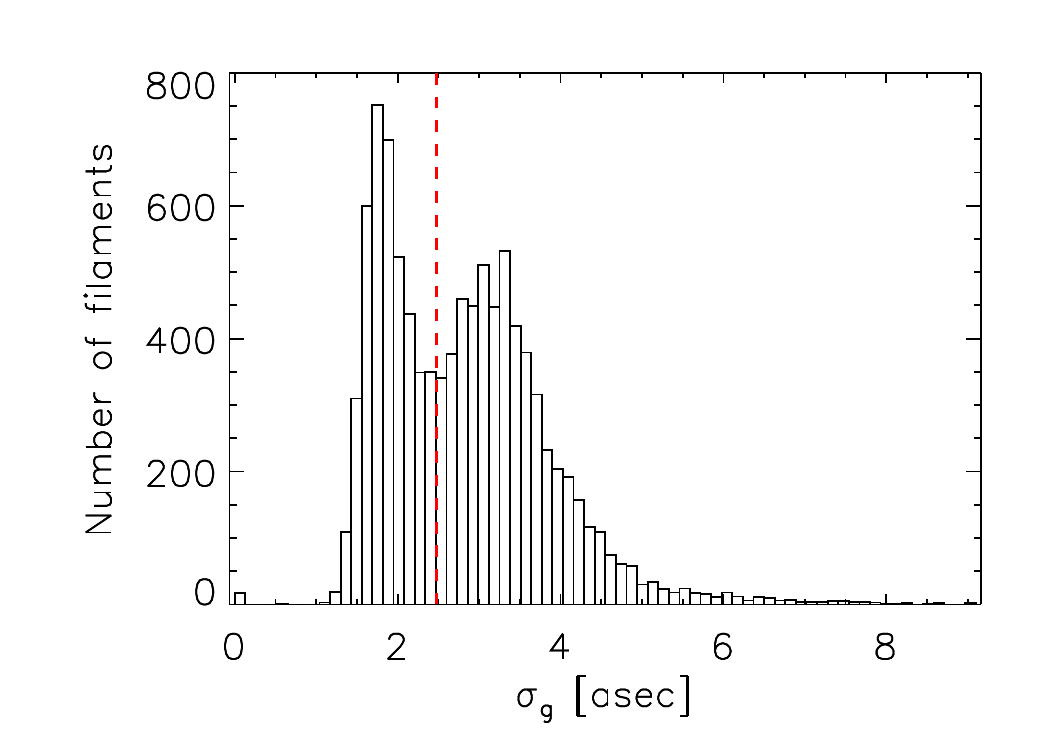} 
   \caption{Histogram of filament widths. The red dashed line (at 2.25 pix $= 2.475''$) indicates an adopted
   separation between likely real filaments and artificially narrow artifacts. {The Gaussian smoothing
   applied to the data ($\sigma_1 = 2.75''$, Section \ref{sec:data}) leads to a smoothed beam with 
   $\sigma = 3.23''$ in the filtered image.}}
   \label{fig:width_hist}
\end{figure}

The median filament widths (i.e. $\sigma_{\rm g}$) exhibits a bimodal distribution,
as shown in Fig.\ \ref{fig:width_hist}.
The minimum at $\sim 2.475''$ appears to serve as a good
first cut at separating real (wider) features, from potentially spurious features 
that have widths that are artificially small relative to the beam.

For the study of NRFs, additional selections were made based
on the filament length, $L$. A constraint of $L > 66''$ was chosen to 
separate NRFs from other real features, e.g. SNRs and 
thermal structures
in the CMZ. A more restrictive constraint of $L > 132''$, provides a
smaller sample, but one that includes very few objects that are not
NRFs {\citep{Yusef-Zadeh:2022a}}.

\begin{figure*}[t]
   \centering
   \includegraphics[height=7in, angle=90]{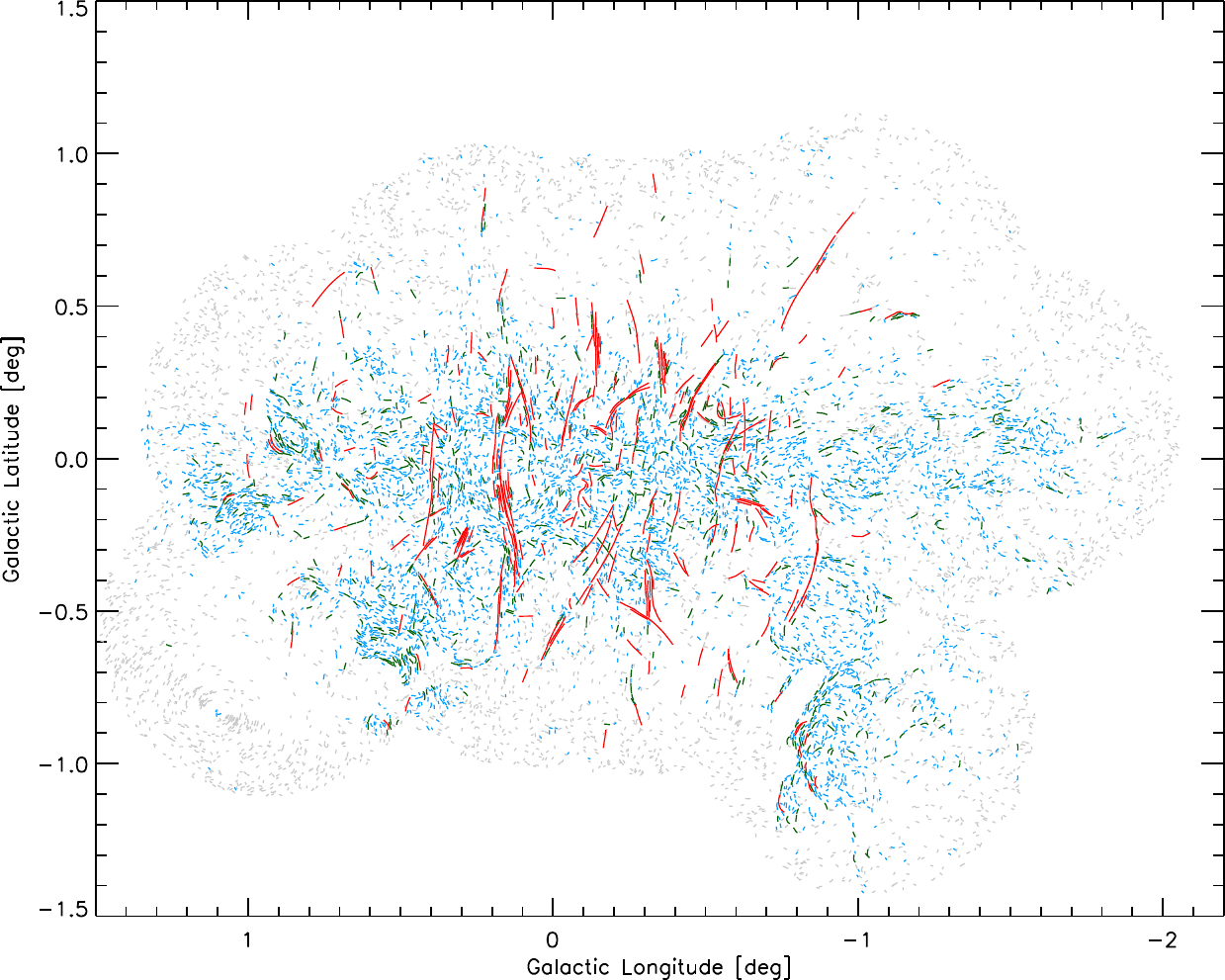} 
   \caption{Map of filaments with selection criteria indicated by color. Gray indicates 
   width $\sigma_{\rm g} < 2.475''$. Wider filaments are colored blue for lengths $33'' < L < 66''$, 
   green for $66'' < L < 132''$, and red for $132'' < L$.}
   \label{fig:selections}
\end{figure*}

\begin{figure*}[t]
   \centering
   \includegraphics[height=7in, angle=90]{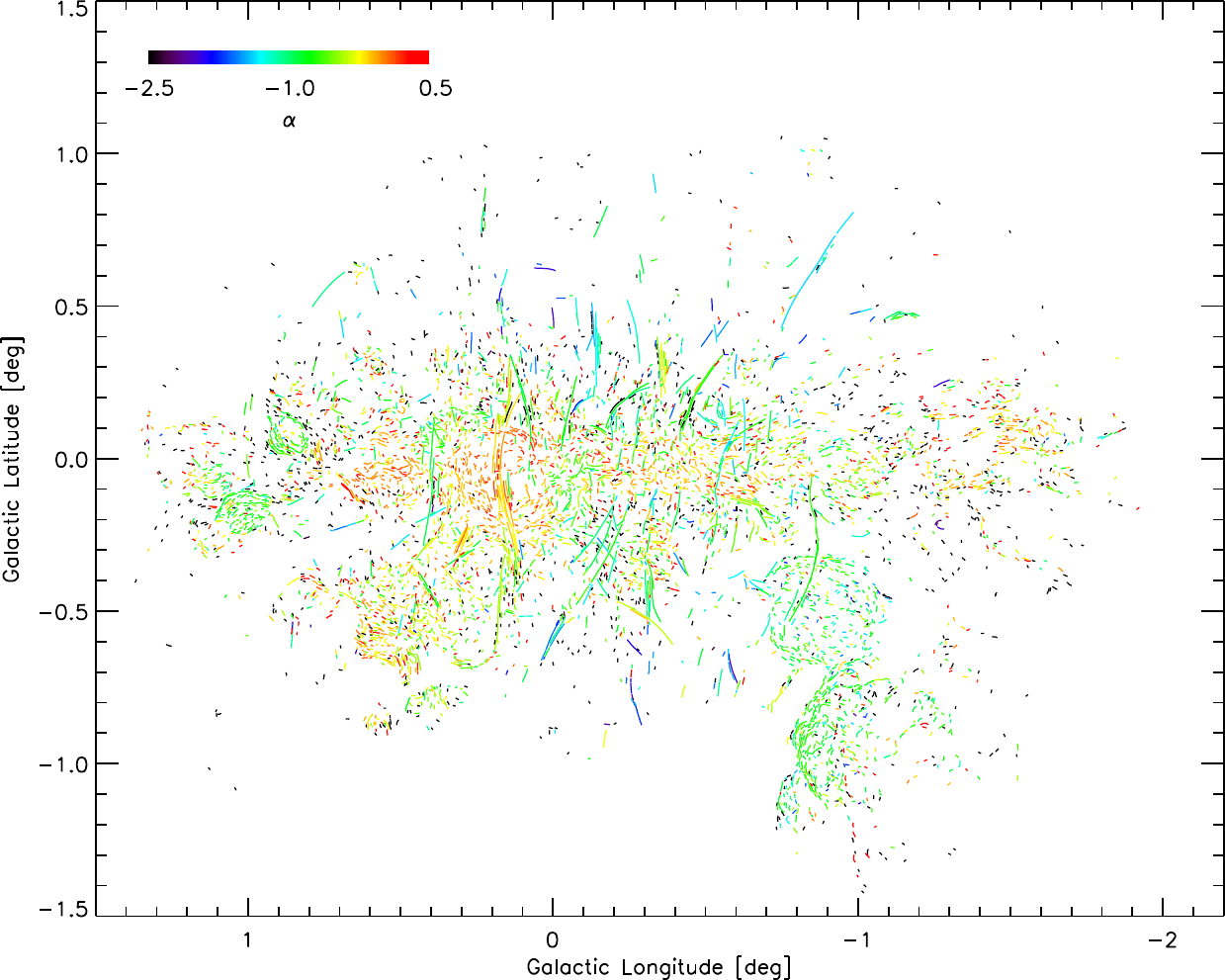} 
   \caption{Map of filament spectral indices. Thermal emission with $\alpha \gtrsim -0.1$ (yellow to red) appears 
   mostly as short filaments with little organization at $|b| < 0.2\arcdeg$. Supernova remnants 
   with $\alpha \approx -0.7$ (green) appear as rounded clusters of curved filaments. 
   The nonthermal radio filaments are more linear, oriented vertically, and generally have somewhat
   steeper spectral indices (blue).}
   \label{fig:alphas}
\end{figure*}

\begin{figure*}[t]
   \centering
   \includegraphics[height=7in, angle=90]{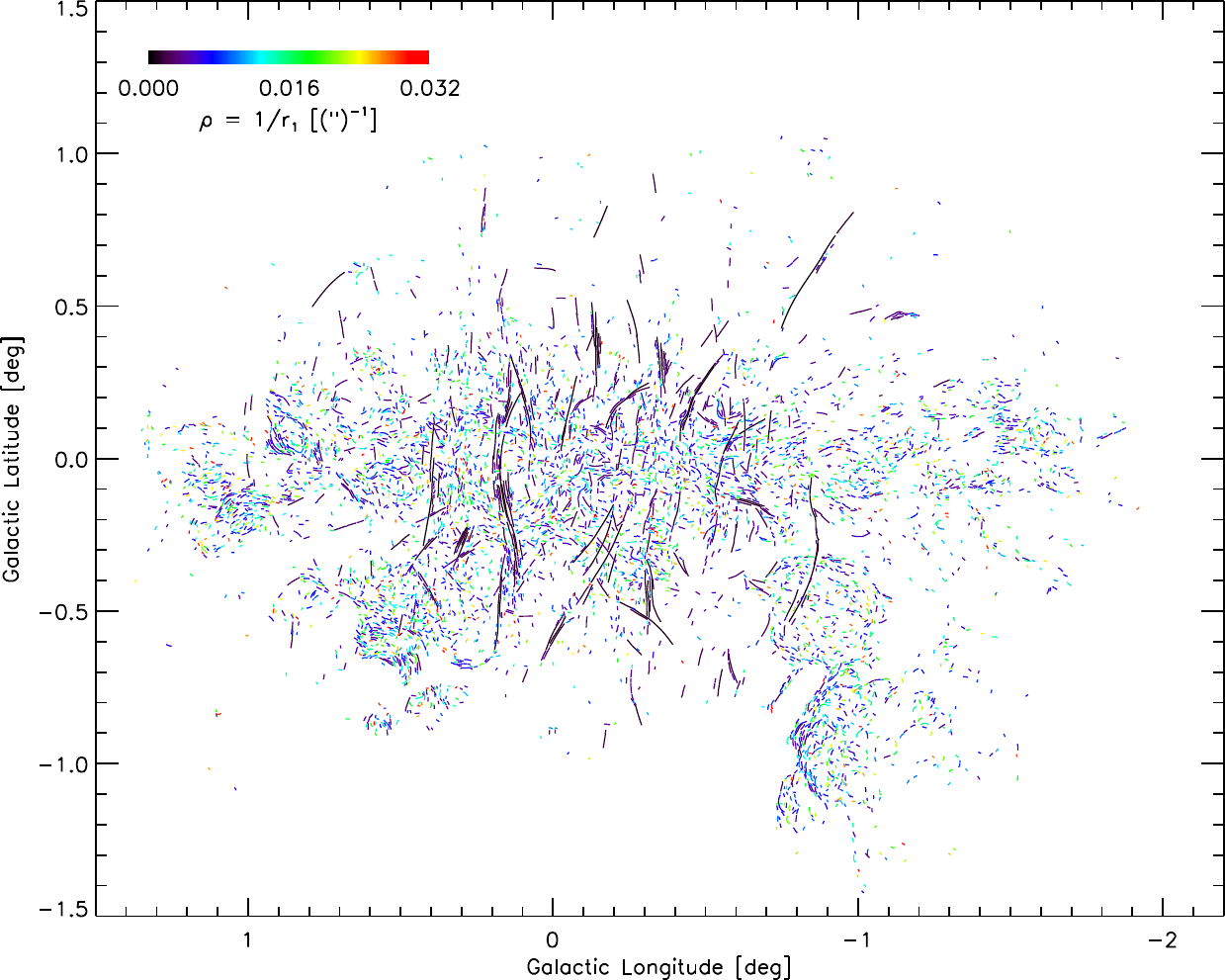} 
   \caption{Map of filament curvatures. Many of the straightest filaments ($\rho \approx 0$)    
   can be recognized as long nonthermal radio filaments here.}
   \label{fig:curvatures}
\end{figure*}

\begin{figure*}[t]
   \centering
   \includegraphics[width=7in, angle=0]{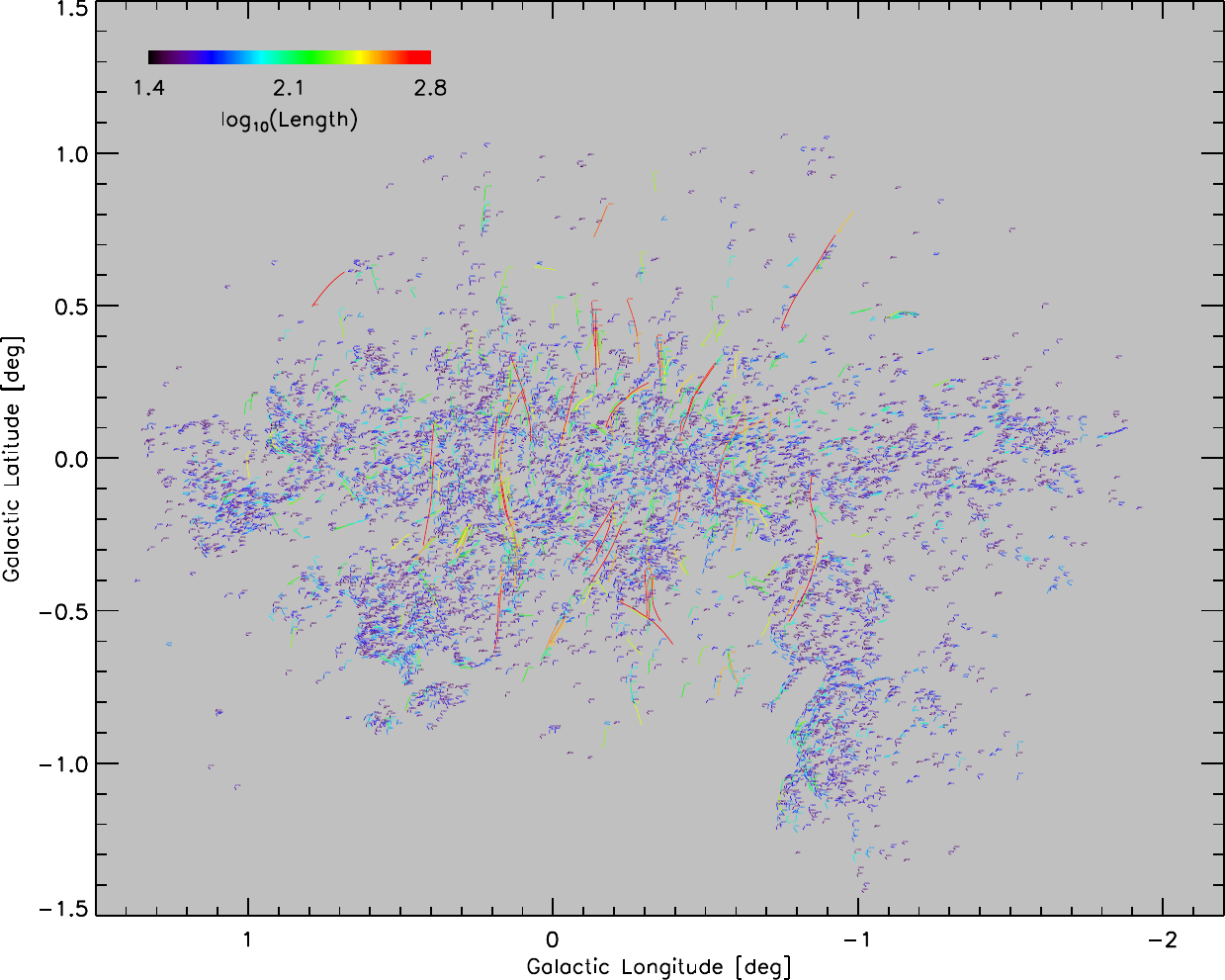} 
   \caption{Finder chart for catalog filaments having widths $\sigma_{\rm g} > 2.475''$. Each filament is labeled with its catalog number, but the numbers
   are very small to minimize confusion and the PDF-format figure must be 
   magnified to read them. The filaments and numbers are color coded by 
   filament length (in arc seconds) to also help filament identification.}
   \label{fig:finder}
\end{figure*}

Figure \ref{fig:selections} illustrates the application of these various 
selection criteria to the cataloged filaments. For various purposes, 
constraints on the spectral index, and curvature may also provide 
useful selection criteria, as illustrated in Figs.\ \ref{fig:alphas} and 
\ref{fig:curvatures}. Figure \ref{fig:finder} serves as finder chart 
to identify specific filaments, but must be magnified for the catalog numbers 
to be legible.

\section{Discussion} \label{sec:discussion}

Because the input image is filtered (twice), small scale detail can be missed. 
{Narrowly separated} filaments in the 
original image can be blended into a single detected filament. Faint details may be unrecognizably 
smoothed into the background. In complex regions, the negative side lobes of bright filaments
or other features may obscure fainter filaments. 

In areas that seem to be covered with an unresolved confusion of faint filaments, only a few 
short loops are typically identified. It is unclear if the emission in these regions is truly diffuse 
or flocculent, or if higher resolution and sensitivity would resolve narrow individual filaments.

When filaments cross a bright point source or extended structure, they may be detected as 
separate objects because the interruption is larger than the gap length. Filaments that cross other 
brighter filaments are also broken into separate segments, because the prior subtraction of the 
detected brighter filament again causes a gap that is larger than the $n_{\rm gap}$ parameter.

Like point source extraction procedures, the automated loop detection falls in completeness
as the filament brightness decreases. 
Thus there are some cases where faint filaments are 
visible to the eye, but have not been detected by the procedure. Assessing the completeness 
by inserting artificial sources into the image and rerunning the source detection procedure
(as with extragalactic source counts) is impractical here because the sources have both a brightness
and a length (unlike point sources) and because the underlying background emission varies 
dramatically across the image. {However, Fig. 9a of \cite{Yusef-Zadeh:2022a} shows a power law fit to 
a histogram filament median intensities ($I_\nu$ from the main catalog). This allows estimates of 
the completeness as $[0.80,\ 0.42,\ 0.10]$ at $I_\nu = [7.6,\ 5.0,\ 3.3]\times 10^{-5}$ Jy beam$^{-1}$,
assuming the reduced numbers of fainter filaments is not due to an intrinsic change in nature of the 
filaments that happens to occur very close to the sensitivity limit.

There is no comparable completeness issue with respect to length, $L$, which is not a factor
in the detectability of a filament. However, note that the parameter $l_{\rm min}$ in Table 1
means that the catalog will not contain any filaments with $L < 33''$. Because the fraction
of NRFs compared to other types of features decreases as $L$ decreases, extending the 
catalog to shorter filaments by setting a smaller value of $l_{\rm min}$ is of
limited value for the study of NRFs, although such a change may be useful in 
alternative cases, such as the study of SNRs.

A} large number of the shorter and fainter loops detected appear to be noise features.
As these noise loops are usually short, they can have small radii of curvature. All loops were
restricted to have $r_{\rm min}>30$ pixels, but a loop of length $L$ must have a radius of curvature 
$r > L/(2\pi)$. Thus, by construction, any loop with $L>188$ pixel must have $r>30$ pixels.

Running the procedure with $n_{\rm gap}=3$ produces generally very similar results for the brighter
filaments, but tends to find different ``noise'' loops. Thus, detection at both $n_{\rm gap}=2$ and 
$n_{\rm gap}=3$ appears to be a good indicator of real features. 
However a general implementation of such a test is difficult because there is not always a 
clear one-to-one match between filaments found in each variation of the processing.

Many features in the brighter SNRs (and some star forming regions) are often detected as relatively 
short and sharply curved loops. Spectral index can also serve in distinguishing
these objects.

Residual side lobe-like patterns can leave concentric rings around 
bright point sources. Portions of these rings are sometimes misidentified 
as loops, though they can mostly be excluded by 
applying the $\sigma_{\rm g} < 2.475''$ constraint.

The filtering and control parameters for {\tt looptracing\_auto4} were set to provide 
reasonable results over a wide area. If one is interested in a specific filament or cluster 
of filaments, it may be useful to modify the filtering or parameters to provide better results
for a particular region of interest. Filamentary structures in several SNRs and star-forming 
regions are part of this catalog, but may also be better defined by re-running 
{\tt looptracing\_auto4} with alternate parameter values.


\begin{acknowledgments}
Work by R.G.A. was supported by NASA under award numbers 80GSFC21M0002 and 
80GSFC24M0006. 
This work is also partially supported by the grants AST-2305857 from the NSF.
{We thank the anonymous referee for their help in improving the manuscript.}
\end{acknowledgments}

\facilities{MeerKAT}

\software{IDLASTRO \citep{Landsman:1995}, {\tt looptracing\_auto4} \citep{Aschwanden:2010}}

\clearpage

\appendix
\section{Curvature Measurements}
\begin{figure}[t]
   \centering
   \includegraphics[width=3in]{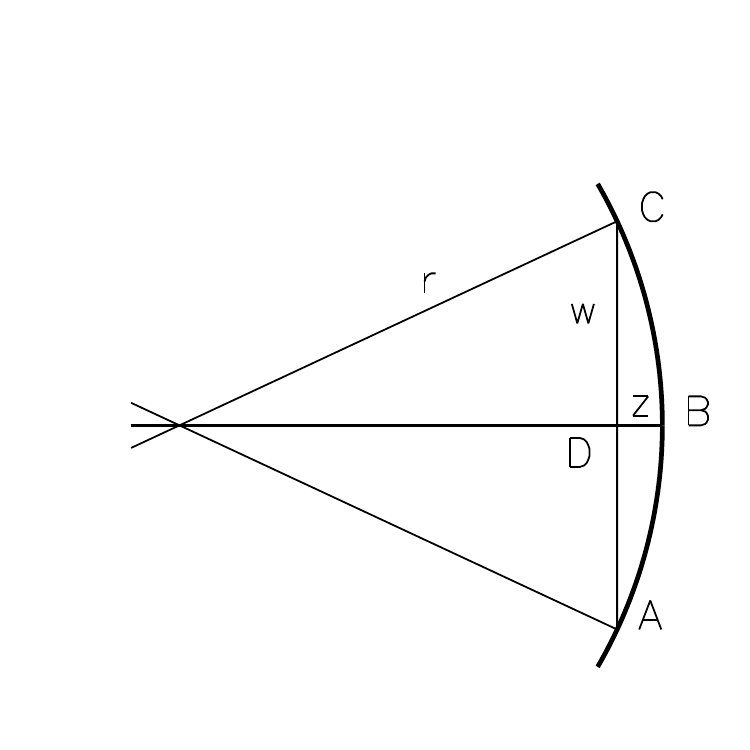} 
   \includegraphics[width=3in]{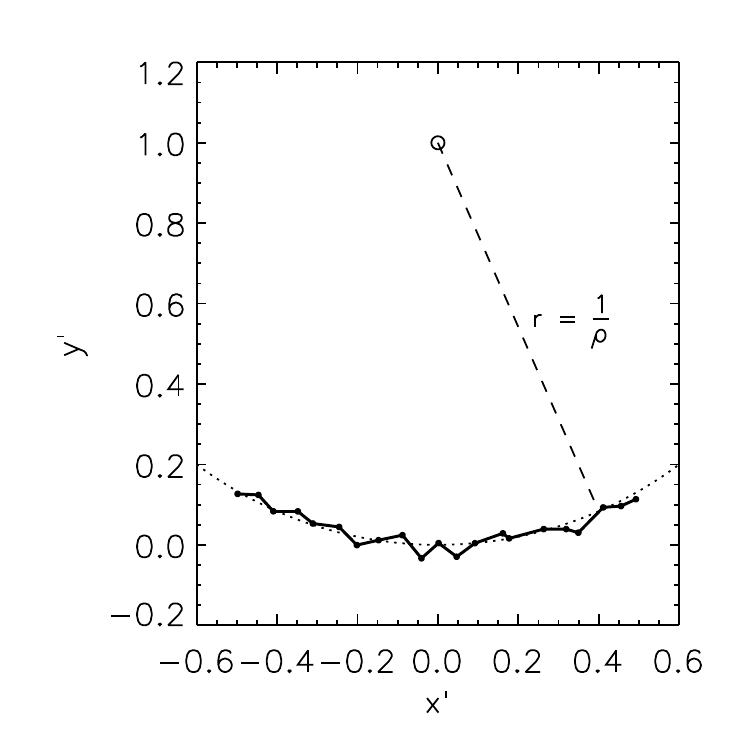} 
   \caption{({a, left}) Schematic curvature measurement at point B on a filament, bracketed by points $A$ and $C$.
   The curvature, $r$, is a simple function of $w$, the half-length of segment $\overline{AC}$, and $z$, the separation 
   of B from the midpoint D of $\overline{AC}$. See Eq.\ \ref{eq:curve1}.
   ({b, right}) Schematic curvature measurement for fitting a circular arc (dotted line) with radius $r$ to
   a filament (connected dots) that has been rotated into a horizontal orientation. The algorithm actually 
   fits for $\rho \equiv 1/r$ since the relevant parameter range is then finite.}
   \label{fig:curve1}
\end{figure}

Local curvature, $r$, as calculated in the main catalog is measured as schematically shown in Fig.\ \ref{fig:curve1}a 
using 3 points $A$, $B$, $C$ as
\begin{equation}
r = (w^2+z^2)/2z
\label{eq:curve1}
\end{equation}
where $w = 0.5*\overline{AC}$ and $z = \overline{BD}$ where $D$ is the midpoint of $\overline{AC}$.
This assumes that the lengths $\overline{AB} = \overline{BC}$.
As applied here, $r$ is calculated for each point $B$ along a loop choosing $A$ and $C$ to be 
7 pixels away along the loop. This applies some smoothing to the curvature measurement. 
Using points closer together might reduce the radius of curvature at some locations, but would 
be more subject to numerical roundoff errors and $r>30$ pixels is already enforced by the 
loop detection algorithm.

The summary catalog lists three measurements of curvature. The first, $r_1$, is this three-point method
applied globally to each filament by using the midpoint for $B$ and the filament endpoints for $A$ and $C$.
The second method, $r_2$, uses a least-squares fit to all of the $\{x,y\}$ locations for each filament 
to solve for the best-fitting radius of curvature. The filament is rotated by the mean position 
angle to $\{x',y'\}$ coordinates in which the long dimension is along the $x'$ axis and the filament
center is at the origin. A circular arc fitting the filament can then be defined by 
\begin{equation}
r^2 = x'^2 + (r-y')^2
\end{equation}
as shown schematically in Fig. \ref{fig:curve1}b, with the center of the circle constrained to 
be at $\{x',y'\}  = \{0,r\}$.
If we substitute $\rho\equiv r^{-1}$ so that the open range of $|r| > 30$ is transformed to 
the closed range of $-0.0333 < \rho < 0.0333$, then we determine $\rho$ by minimizing $\chi^2$
for the model
\begin{equation}
y' = \frac{1}{\rho} [1\pm(1-x'^2\rho^2)^{0.5}].
\end{equation}
We report the radius of $r = |\rho^{-1}|$, as the sign of $\rho$ only indicates whether 
the filament is arced towards $+y'$ or $-y'$.
The third
curvature, $r_3$, listed in the summary catalog is the median of the local three-point curvature radii in the main 
catalog for each filament.

\bibliography{looptracing}{}
\bibliographystyle{aasjournalv7}

\end{document}